\documentclass[prl,twocolumn,showpacs,amsmath,amssymb]{revtex4}
\usepackage{graphicx}%
\usepackage{psfrag}  %
\usepackage{hyperref} %
\hypersetup{pdfnewwindow=true, colorlinks=true, %
 linkcolor=blue, anchorcolor=blue,
 citecolor=blue, filecolor=blue,
 menucolor=blue, urlcolor=blue}
\begin{document}

\title{Electric polarization in a Chern insulator}

\author{Sinisa Coh} 
\email{sinisa@physics.rutgers.edu} 
\author{David Vanderbilt} 
\affiliation{
  Department of Physics \& Astronomy, Rutgers University, Piscataway,
  NJ 08854-8019, USA
}

\date{\today}

\def\mlab#1{}
\def\a{{\bf a}}
\def\b{{\bf b}}
\def\P{{\bf P}}
\def\k{{\bf k}}
\def\R{{\bf R}}
\def\r{{\bf r}}
\def\j{{\bf j}}
\def\J{{\bf J}}
\def\n{{\bf n}}
\def\A{{\bf A}}
\def\E{{\bf E}}
\def\ks{{k_2^{\star}}}
\def\ks{{k_2^*}}
\def\kc{{k_2^{\times}}}
\def\S{{\cal A}}
\def\S{S}
\def\ket#1{\vert #1 \rangle}
\def\bra#1{\langle #1 \vert}
\def\me#1#2#3{\bra{#1}#2\ket{#3}}
\def\ti{t_{\rm i}}
\def\tf{t_{\rm f}}
\def\ai{\alpha_{\rm i}}
\def\af{\alpha_{\rm f}}
\def\supi{^{({\rm i})}}
\def\supf{^{({\rm f})}}
\def\finP1{\mathcal{P}_1}
\def\finPvec{\boldsymbol{\mathcal{P}}}

\begin{abstract}  
  We extend the Berry-phase concept of polarization to
  insulators having a non-zero value of the Chern invariant.
  The generalization to such Chern insulators requires special
  care because of the partial occupation of chiral edge states.
  We show how the integrated bulk current arising from an adiabatic
  evolution can be related to a difference of bulk polarizations.
  We also show how the surface charge can be related to the bulk
  polarization, but only with a knowledge of the wavevector
  at which the occupancy of the edge state is discontinuous.
  Furthermore we present numerical calculations on a model Hamiltonian
  to provide additional support for our analytic arguments.
\end{abstract}


\maketitle

In 1988 Haldane pointed out that an insulating
crystal with broken time-reversal symmetry may exhibit a quantized
Hall conductance even in the absence of a macroscopic magnetic field
\cite{haldane-prl88}.  We shall refer to such a material as a ``Chern
insulator'' (CI) because it necessarily would have a non-zero Chern
invariant associated with its manifold of occupied Bloch states
\cite{thouless,tknn}.  While no CI has yet been discovered
experimentally, there appears to be no reason why one could
not exist, and theoretical models that behave as CIs
are not difficult to construct.  It seems plausible that the
current blossoming of interest in exotic non-collinear magnets
and multiferroics
could yield an experimental example before long.

CIs occupy a middle ground between metals and
ordinary insulators.  Like metals, their conductivity tensor
$\sigma_{\alpha\beta}$ is non-zero, their surfaces are metallic
(as a result of topological edge states crossing the Fermi energy),
and it is impossible to construct exponentially
localized Wannier functions (WFs) for them \cite{brouder-prl07}.
On the other hand, only the off-diagonal (dissipationless)
elements of $\sigma_{\alpha\beta}$ can be non-zero, the chiral
edge states decay exponentially into the bulk,
the one-particle density matrix decays
exponentially in the interior \cite{thonhauser-prb06},
and the localization measure $\Omega_{\rm I}$ \cite{resta-prl99,marzari-prb97}
is finite \cite{thonhauser-prb06} as in other insulators.
Overall it appears natural to regard a CI as an unusual species
of insulator, but many aspects of its behavior remain open to
investigation.

As is well known, the electric polarization $\bf P$
is not well-defined in a metal.  For an
ordinary insulator, its definition alternatively in terms of Berry
phases or WFs is by now well
established~\cite{kingsmith-prb93,vanderbilt-prb93,
resta-chapter06}.  For a CI, the
absence of a Wannier representation removes the possibility of using
it to define the polarization, and we shall show below that there is a
fundamental difficulty with the Berry-phase definition as well.  In
view of the presence of dissipationless currents and metallic edge
states, one might be tempted to conclude that $\bf P$ is not
well-defined at all in a CI.  On the other hand, $\Omega_{\rm I}$ is
related to the fluctuations of $\bf P$ \cite{souza-prb00},
and the finiteness of this
quantity~\cite{thonhauser-prb06} suggests that the polarization
might be well-defined after all.

The purpose of this Letter is to discuss whether, and in what sense, a
definition of electric polarization is possible in a CI.
We demonstrate that the usual Berry-phase definition
does remain viable if it is interpreted with care when
connecting it to observables such as the internal current that flows in
response to an adiabatic change of the crystal Hamiltonian,
or to the surface charge at the edge of a bounded sample.

For the remainder of this Letter we restrict ourselves to the case
of a 2D crystalline insulator having a single isolated
occupied band.  The generalization to the case of a
3D multiband
insulator is not difficult, but would complicate the
presentation.  We also restrict ourselves to a
single-particle Hamiltonian, noting that the principal difficulties in
understanding CIs occur already at the one-particle
level.  The lattice vectors $\a_1$ and $\a_2$ are related
to the reciprocal lattice vectors $\b_1$ and $\b_2$ in the
usual way ($\b_i\cdot\a_j=2\pi\delta_{ij}$) and the cell area
is $\S=|\a_1\times\a_2|$.

The Berry-phase expression for the electric polarization can be
written as
\mlab{eq:new}
\begin{equation}
\P_{[\k_0]}=\frac{e}{(2\pi)^2} \,{\rm Im} \int_{[\k_0]} d\k\,
    \me{u_\k}{\nabla_\k}{u_\k}
\label{eq:new}
\end{equation}
where $e$ is the charge quantum ($e>0$),
$\ket{u_\k}$ are the cell-periodic Bloch functions, and
$[\k_0]$ indicates the parallelogram reciprocal-space unit cell
with origin at $\k_0$ (that is, with vertices $\k_0$, $\k_0+\b_1$,
$\k_0+\b_1+\b_2$, and $\k_0+\b_2$).  In an ordinary insulator
one insists on a smooth and periodic choice of gauge
(relative phases of the $\ket{u_\k}$)
in Eq.~(\ref{eq:new}), and $\P$ is well defined (modulo $e\R/S$,
where $\R$ is a lattice vector \cite{kingsmith-prb93})
independent of $\k_0$.  However, in a
CI such a gauge choice is no longer possible.
To see this, we decompose $\P_{[\k_0]}=P_1\a_1+P_2\a_2$,
$\k=k_1\b_1+k_2\b_2$, and $\k_0=\kappa_1\b_1+\kappa_2\b_2$, and
rewrite Eq.~(\ref{eq:new}) as
\mlab{eq:polBp}
\begin{equation}
  P_1^{[\kappa_2]} = \frac{-e}{\S} \int_{\kappa_2}^{\kappa_2+1} dk_2 \,
 \frac{\theta_1(k_2)}{2\pi} \,,
\label{eq:polBp}
\end{equation}
\mlab{eq:polAv}
\begin{equation}
  \theta_1(k_2) = -{\rm Im} \int_{\kappa_1}^{\kappa_1+1} dk_1
   \me{u_{k_1,k_2}} {\partial_{k_1}} {u_{k_1,k_2}} \,.
  \label{eq:polAv}
\end{equation}
Eq.~(\ref{eq:polAv}) is a Berry phase and is gauge independent modulo
$2\pi$ (independent of $\kappa_1$).
This allows us to make an arbitrary choice
of branch for $\theta_1(k_2=\kappa_2)$ and to insist,
as part of the definition of $P_1^{[\kappa_2]}$,
that $\theta_1(k_2)$ should remain continuous as $k_2$ is increased from
$\kappa_2$ to $\kappa_2+1$.
Since states at $(k_1,\kappa_2)$ and $(k_1,\kappa_2+1)$ are equivalent,
it follows that
\mlab{eq:per}
\begin{equation}
\theta_1\Big\vert_{\kappa_2}^{\kappa_2+1} = -2 \pi C
\label{eq:per}
\end{equation}
where $C$ is an integer.  In fact $C$ just defines the Chern number,
and the insulator is a CI if $C\ne0$.
For simplicity we focus henceforth on a CI having $C=\pm1$.

\begin{figure}[!t]
  \includegraphics{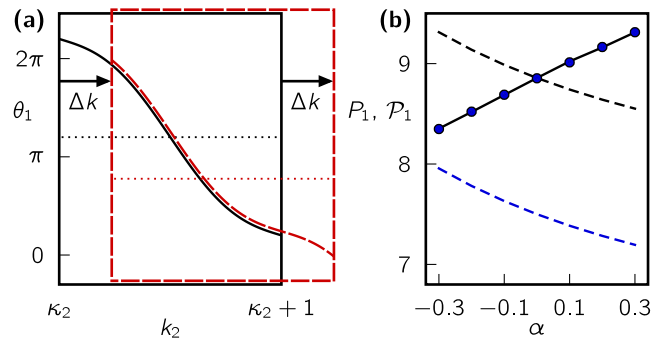}
  \caption{\label{fig:phnu} (Color online) (a) Sketch of
    $\theta_1(k_2)$ in a Chern insulator ($C=+1$). Solid black and
    dashed red frames indicate reciprocal-cell origin chosen at
    $\kappa_2$ and $\kappa_2+\Delta k$ respectively.  Dotted lines
    indicate corresponding averages, proportional to $P_1$.  (b)
    Computed $P_1(\alpha)$ and $\finP1(\alpha)$
    for the modified Haldane model, in units
    of $-0.01e/S$, for adiabatic (dashed lines) and thermal (solid
    line and symbols) filling.  See text.}
\end{figure}

Using Eqs.~(\ref{eq:polBp}-\ref{eq:polAv}) and similar equations
for $P_2$, we have arrived at a definition $\P_{[\k_0]}$ that is
well-defined, modulo $e\R/S$ as usual, even for a CI.
However, as illustrated in Fig.~\ref{fig:phnu}(a),
\mlab{eq:k0dep}
\begin{equation}
\P_{[\k_0+\Delta\k]}=\P_{[\k_0]}-\frac{eC}{2\pi}\,\hat{\bf z}\times\Delta\k
\label{eq:k0dep}
\end{equation}
where $\hat{\bf z}$ is the unit vector along $\a_1\times\a_2$.
This dependence on $\k_0$ clearly presents a problem for the interpretation
of Eq.~(\ref{eq:polBp}) as a ``physical'' polarization
in the case of a CI.

However, let us recall how the concept of polarization is
{\it used}.  For a normal insulator at least \cite{kingsmith-prb93},
the change of
polarization during an adiabatic change of some internal parameter
of the system from time $\ti$ to $\tf$ is given by
\mlab{eq:prA}
\begin{equation}
  \int_{\ti}^{\tf} dt \, \J(t)
  =\P_{[\k_0]}\supf-\P_{[\k_0]}\supi \quad \mathrm{(modulo}\;e\R/S) \,,
\label{eq:prA}
\end{equation}
where $\J(t)$ is the cell-averaged adiabatic current flowing in
the bulk.
A related statement, connected with the requirement that the charge 
pumped to the surface must be consistent with Eq.~(\ref{eq:prA}),
is that the charge on an insulating surface normal to reciprocal
vector $\b_1$ is \cite{vanderbilt-prb93}
\mlab{eq:prB}
\begin{equation}
  \sigma = \P\cdot\hat{\b}_1 \quad \mathrm{(modulo}\;e/a_2)\,.
\label{eq:prB}
\end{equation}
Eqs.~(\ref{eq:prA}) and (\ref{eq:prB}) embody the attributes of a
useful definition of $\bf P$.
In the remainder of this Letter, we demonstrate that
a generalized definition of $\bf P$, having similar attributes,
can be given in a CI.  We first show that Eq.~(\ref{eq:prA})
remains correct, provided that
the {\it same} $\k_0$ (i.e., the same reciprocal-space cell) is
used for $\P\supi$ and $\P\supf$ in Eq.~(\ref{eq:prA}).
We also show that Eq.~(\ref{eq:prB}) must be modified and
explain how.  We provide numerical tests as well as
analytic arguments for both claims.

We begin by giving two arguments for the correctness of
Eq.~(\ref{eq:prA}) in the CI case.
First, it is straightforward to see that the the contribution
to $J_1(t)$ can be computed independently for each $k_2$ 
\cite{vanderbilt-prb93}, with
the problem in $(k_1,t)$ space effectively corresponding to that
of an ordinary 1D crystal.  Thus, the derivation
of Eq.~(\ref{eq:prA}) given in Ref.~\cite{kingsmith-prb93} goes
through unchanged for the CI case.
Second, we note that the expected result is obtained for the
special case that the parameter of interest is a spatially
uniform but time-dependent vector potential $\A(t)$.  Since
a slow turning on of $\A(t)$ causes state $u_\k e^{i \k\cdot\r}$ 
to evolve into $u_{\k+(e/\hbar c)\A} e^{i \k\cdot\r}$,
it follows that
\mlab{eq:Adep}
\begin{equation}
\P_{[\k_0]}^{[\A]}=\P_{[\k_0]}^{[\A=0]}-\frac{e^2C}{hc}
    \,\hat{\bf z}\times\A \,.
\label{eq:Adep}
\end{equation}
But a time varying vector potential generates an electric field
$\E=(-1/c)d\A/dt$, so that
$\J=(C e^2/h)\hat{\bf z}\times{\E}$.
The transverse conductivity $\sigma_{xy}$ is thus quantized in
units of $e^2/h$, expressing the fact
that a CI is a realization of the integer quantum
Hall effect \cite{haldane-prl88}.

We further confirm the validity of Eq.~(\ref{eq:prA}) by
numerically testing our prediction on the Haldane model
[\onlinecite{haldane-prl88}],
a tight-binding model for spinless electrons
on a honeycomb lattice at half filling with
staggered site energies and complex second-neighbor hoppings
chosen so that $C$=1.
Using the notation of Ref.~\cite{haldane-prl88}, we adopt
parameters $t_1=1$, $t_2=1/3$, $\phi=\pi/4$, 
$\Delta=2/3$ and the lattice vectors
$\a_1=a_0(\sqrt{3}\hat{x}+\hat{y})/2$ and $\a_2=a_0\hat{y}$
(so that $a_1=a_2=a_0$).
Furthermore, we modify the first-neighbor hopping
$t_1\rightarrow t_1(1+\alpha)$ on the bonds parallel to
$\a_1+\a_2$ so as to break the threefold rotational symmetry
and allow an adiabatic current to flow as $\alpha$ is
varied.
The compensating ionic
charge is assumed to sit on the site with lower site energy.

\begin{figure}[!t]
  \includegraphics{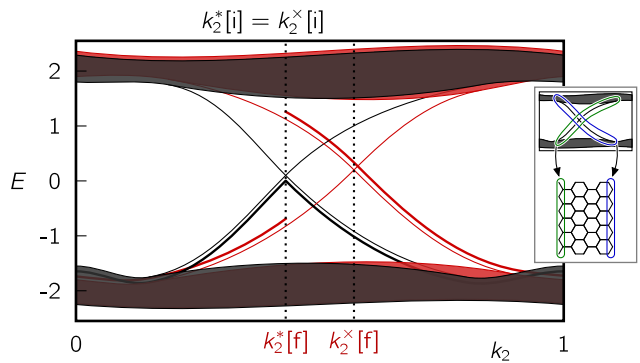}
  \caption{\label{fig:edg} (Color online) Sketch of a 
    band structure of a finite
    ribbon of a Chern insulator. Solid regions indicate projected bulk
    bands; thin solid lines are edge states.  Black and red
    correspond to $\alpha=\ai$ and $\alpha=\af$ respectively;
    corresponding values of $\kc$ are indicated.
    Thick lines indicate filling of edge states as dictated by
    $\ks$, chosen to illustrate system thermalized at $\ai$ and
    then carried adiabatically to $\af$.  Inset: edge states
    associated with left (green) and right (blue) surfaces.}
\end{figure}

We consider an infinite strip of the Haldane model $N_1$ cells wide
and extending to $\pm\infty$ along $y$, as sketched in the inset of
Fig.~\ref{fig:edg}.  States $\psi_{nk_2}(\r)$ are labeled by $k_2$,
which remains a good quantum number, and an additional index
$n=1,...,2N_1$.  The dipole moment across the strip, per unit length,
is
\mlab{eq:fin}
\begin{equation}
  \finP1=\frac{-e}{N_1\S}\int_0^{1} dk_2 \sum_{n\in {\cal N}(k_2)}
  \langle \psi_{nk_2} \vert r_1 \vert \psi_{nk_2} \rangle \,,
  \label{eq:fin}
\end{equation}
where position vector $\r$ is decomposed as $\r=r_1\a_1+r_2\a_2$
and ${\cal N}(k_2)$ is the set of occupied states 
to be discussed shortly. In the limit of large $N_1$,
we associate the integrated current that flows along $\hat{x}$
in the interior of the strip during an adiabatic evolution from
$\alpha=\ai$ to $\alpha=\af$ with the
corresponding change in $\finP1$, since by continuity the
charge must arrive at the surface.  We then compare this with the
change of $P_1$ evaluated using a single bulk unit cell via
Eqs.~(\ref{eq:polBp}-\ref{eq:polAv}) to validate the theory.

There is a subtlety, however.  Neutrality implies that
${\cal N}(k_2)$ contains $N_1$ states,
but which ones?  The problem arises because
a CI is topologically required to have chiral
metallic edge states.  Our ribbon of CI
therefore has one band of edge states along its left (L) edge
and one along its right (R) edge (see inset of Fig.~\ref{fig:edg}).
For any given $\alpha$, let $\kc(\alpha)$ be the value of $k_2$ at
which L-edge and R-edge bands cross.
A {\it thermalized} filling of the edge states would correspond
to the thick black curve for case $\ai$ in Fig.~\ref{fig:edg},
where the $N_1$ lowest-energy states are occupied at each $k_2$
and $\epsilon_{\rm F}=\epsilon({\kc})$.  Defining $\ks$  to be the
point at which the occupation switches between L and R edge states,
we have $\ks=\kc$ for the thermalized case.

In general $\kc(\alpha)$ varies with $\alpha$.  However, $\ks$
cannot change during an {\it adiabatic} evolution.  Because we want to
``measure'' the polarization by the charge that accumulates at the
surface, we specify that the adiabatic evolution is fast compared
to the tunneling time between edge states but slow compared to all
other processes, so that electrons cannot scatter between
edges.  Thus if we thermalize the system at $\ai$ and then
adiabatically carry the system from $\ai$ to $\af$, we arrive at
the {\it adiabatic} filling illustrated by the thick red curve for
case $\af$ in Fig.~\ref{fig:edg}.

We thus expect that the change in polarization calculated from the
right-hand side of Eq.~(\ref{eq:prA}) from the bulk bandstructure
using Eqs.~(\ref{eq:new}-\ref{eq:polAv})
should match that given by the change of Eq.~(\ref{eq:fin}) {\it only}
if the adiabatic filling is maintained.  We have confirmed this
numerically for our modified Haldane model.
The polarization as a function of $\alpha$ calculated using
Eq.~(\ref{eq:fin}) and using the right-hand side of Eq.~(\ref{eq:prA}) is
indicated in Fig.~\ref{fig:phnu}(b) with black and blue dashed lines
respectively \cite{footnote1}.
Eqs.~(\ref{eq:polBp}-\ref{eq:polAv}) were evaluated on a
$300\times300$ k-point mesh. Eq.~(\ref{eq:fin}) was calculated
using five values of $N_1\in[25,70]$ and then extrapolating
to infinity, while the $k_2$ integral was discretized with
5000 k-points.
While there is a vertical offset between these curves that depends
on the choice of $\k_0$ in Eq.~(\ref{eq:prA}), the {\it differences}
$\Delta P_1$ between different $\alpha$ are correct at the
level of $10^{-5}$.
On the other hand, the results obtained with the thermalized
filling in Eq.~(\ref{eq:fin}), shown by the solid line
in Fig.~\ref{fig:phnu}(b), are drastically different.  These results
confirm that the appropriate comparison is with the adiabatic filling,
and provide numerical confirmation that Eq.~(\ref{eq:prA}) is indeed
satisfied even in a CI.

\mlab{fig:edg}

We now turn to Eq.~(\ref{eq:prB}).
A naive generalization to the CI case
might be that $\sigma=\P_{[\k_0]}\cdot\hat{\b}_1$
(modulo $e/a_2$), but this cannot be correct.  First, the
left-hand side should be independent of $\k_0$, but the right-hand
side is not.  Second, the usual proof for ordinary insulators
of the connection between surface charge and bulk polarization
assumes that the surface is insulating, with
the Fermi level lying in a gap common to both the bulk and surface
\cite{vanderbilt-prb93}.  When chiral edge states are present,
the surfaces cannot be insulating, so the usual conditions are
violated.

To show how Eq.~(\ref{eq:prB}) can be corrected for the case of a
CI, let us again consider our Haldane-model ribbon at
some fixed $\alpha$. Its surface charge $\sigma$
can be calculated from $\sigma=\finPvec\cdot\hat{\b}_1=(S/a_2)\finP1$
with $\finP1$ evaluated using Eq.~(\ref{eq:fin}), but its value will
depend on the the choice of the $\ks$ at which the occupation of
the edge state has its discontinuity, so that
\mlab{eq:sigone}
\begin{equation}
  \sigma^{[\ks]}=\frac{-e}{N_1a_2}\int_0^1 dk_2
\sum_{n\in {\cal N}}
\langle \psi_{nk_2} \vert r_1 \vert \psi_{nk_2} \rangle \,,
  \label{eq:sigone}
\end{equation}
where $\cal N$ is the set of $N_1$ occupied states at 
$k_2$ given the specified $\ks$ (i.e., the choice
whether the L or R edge state is included in $\cal N$ flips as
$k_2$ passes through $\ks$).

Since the surface charge theorem of Eq.~(\ref{eq:prB}) for ordinary
insulators was demonstrated via the Wannier representation
\cite{vanderbilt-prb93}, we take the same approach here.
However, well-localized bulk WFs do not exist in a CI
\cite{brouder-prl07}, so we focus instead on ``hybrid
Wannier functions'' (HWFs) \cite{sgiarovello}
in which the Fourier transform from
Bloch functions is carried out in the $r_1$ direction only.
Thus $k_2$ remains a good quantum number and the HWF
\begin{equation}
W_{k_2} (r_1,r_2) = \sqrt{N_1}
  \int_0^1 d k_1 \; \varPsi_{k_1k_2} (r_1,r_2)
\end{equation}
is well localized only in the $\a_1$ direction.  Using these
we can represent the polarization
\mlab{eq:hybCen}
\begin{equation}
P_1^{[\kappa_2]}= \frac{-e}{\S} \displaystyle \int_{\kappa_2}^{\kappa_2+1} dk_2 
\;\rho_{k_2}^{[\kappa_2]}
\label{eq:hybCen}
\end{equation}
in terms of the HWF center $\rho_{k_2}^{[\kappa_2]}=
\me{W_{k_2}}{r_1}{W_{k_2}}$.
We require $\rho$ to be a continuous function of
$k_2\in[\kappa_2,\kappa_2+1]$ so as to guarantee a result
that is equivalent to Eqs.~(\ref{eq:new}-\ref{eq:polAv}).

To make the connection between Eqs.~(\ref{eq:sigone}) and (\ref{eq:hybCen}),
we recast the former by constructing Wannier-like functions
along the $\a_1$ direction for the finite-width strip, starting
from the $N_1\times N_1$ matrix
$
  {\cal R}^{[\ks]}_{mn,k_2}= \me{\psi_{mk_2}}{r_1}{\psi_{nk_2}}
$,
where $m,n\in{\cal N}$ as specified by $\ks$.  The $N_1$
eigenvectors of ${\cal R}^{[\ks]}_{k_2}$ correspond to states that
are Bloch-like along $r_2$ but localized along $r_1$, which we
refer to as ribbon HWFs, and the eigenvalues
$\varrho_{jk_2}^{[\ks]}$ locate their centers of charge.
Using the basis-independence of the
trace, Eq.~(\ref{eq:sigone}) can now be rewritten as
\mlab{eq:sigEig}
\begin{equation}
  \sigma^{[\ks]}=\frac{-e}{N_1a_2} \int_0^{1} dk_2
  \sum_j \varrho_{jk_2}^{[\ks]} \,.
  \label{eq:sigEig}
\end{equation}

\begin{figure}[!t]
  \includegraphics{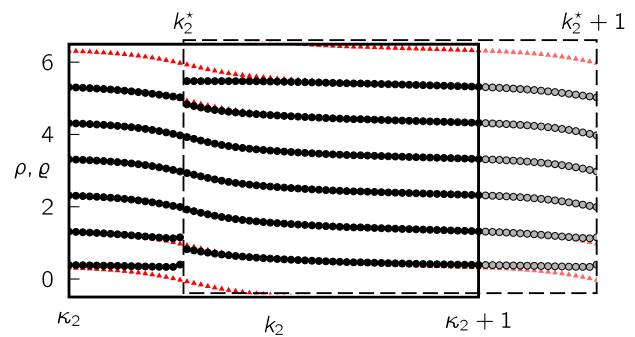}
  \caption{\label{fig:hyb} (Color online) Black dots show ribbon HWF
    centers $\varrho_{jk_2}^{[\ks]}$ and red triangles bulk HWF centers
    $\rho_{k_2}^{[\kappa_2]}$ and its periodic images as a function of
    $k_2$. Dashed frame corresponds to choice of origin at
    discontinuity in $\varrho_{jk_2}^{[\ks]}$, $\ks$. }
\end{figure}

The similarity between Eqs.~(\ref{eq:hybCen}) and (\ref{eq:sigEig})
suggests that these can be connected.  Since $k_2$ is a good quantum number,
each $k_2$ can be treated independently.
For each $k_2$ we can
compare the infinite (bulk) 1D system described by Eq.~(\ref{eq:hybCen})
with the finite (ribbon) 1D system described by Eq.~(\ref{eq:sigEig}).
The essential observation is that, in the limit of large $N_1$, the
HWF centers $\varrho_{jk_2}$ deep inside the ribbon
converge to the bulk $\rho_{k_2}$, modulo an integer
\cite{vanderbilt-prb93}.
This is illustrated in Fig.~(\ref{fig:hyb}), where
both sets of HWF centers are plotted as a function of $k_2$ for
a ribbon of width $N_1=6$.
Furthermore, the fact that the occupation of edge states switches
between L and R edge at $\ks$ is reflected in the discontinuity of
ribbon HWF centers $\varrho_{jk_2}$ at $\ks$. On the other hand, the bulk
HWF centers $\rho_{k_2}$ are chosen to be continuous across $\ks$. We
can account for this discrepancy either by including a correction term
proportional to $(\ks-\kappa_2)$,
\mlab{eq:finAdd}
\begin{equation}
  \sigma^{[\ks]}=\frac{1}{a_2}\left[SP_1^{[\kappa_2]}+eC
    (\ks-\kappa_2) \right] \quad \mathrm{(mod}\;e/a_2)\,,
\label{eq:finAdd}
\end{equation}
or by realizing that by the virtue of Eq.~(\ref{eq:k0dep}) this is
equivalent to shifting the reciprocal space origin to $\ks$,
\mlab{eq:finOr}
\begin{equation}
  \sigma^{[\ks]}=\frac{S}{a_2}P_1^{[\ks]} \quad \mathrm{(mod}\;e/a_2)\,,
\label{eq:finOr}
\end{equation}
as can be seen from the dashed frame in Fig.~\ref{fig:hyb}.
Eq.~(\ref{eq:finAdd}) or (\ref{eq:finOr}) is the appropriate generalization
of the surface charge theorem, Eq.~(\ref{eq:prB}), to the case of
a CI, and should be correct in
large $N_1$ limit for {\it both} thermalized and adiabatic fillings as long
as the appropriate $k_2^{\star}$ is used.

\mlab{fig:hyb}

We have also tested the correctness of this formula using our
numerical calculations on the modified Haldane model.  Recall that
the solid curve in Fig.~\ref{fig:phnu}(b) represents the surface
charge as computed from Eq.~(\ref{eq:fin}) for the thermalized
case.  
For each $\alpha$, we first locate $\kc$ using  
1000 k-points
on a ribbon of width $N_1=70$ and evaluate Eq.~(\ref{eq:finOr})
with $\ks=\kc$
using Eqs.~(\ref{eq:polBp}-\ref{eq:polAv}) on a $250\times250$
k-point mesh. The resulting values are plotted as blue dots in
Fig.~\ref{fig:phnu}(b).  The agreement is excellent.

In summary, we have generalized the Berry-phase concept of polarization
to the case of a Chern insulator.  The integrated current flow
during adiabatic evolution is given by Eq.~(\ref{eq:prA}), where
the reciprocal-space cell must be the same in both
terms on the right-hand side.  The surface charge at an
edge of a bounded sample is given by Eq.~(\ref{eq:finOr}),
where $\ks$ specifies the wavevector at which the
occupation discontinuity occurs in the chiral edge state.
These results may be of use in understanding the physical
properties of these topological insulators, and perhaps in
searching for experimental realizations.

We acknowledge useful discussions with P. Chandra. This work was
supported by NSF Grant DMR-0549198.
\vspace{-0.75cm}
\bibliography{pap}

\begin{thebibliography}{13}
\expandafter\ifx\csname natexlab\endcsname\relax\def\natexlab#1{#1}\fi
\expandafter\ifx\csname bibnamefont\endcsname\relax
  \def\bibnamefont#1{#1}\fi
\expandafter\ifx\csname bibfnamefont\endcsname\relax
  \def\bibfnamefont#1{#1}\fi
\expandafter\ifx\csname citenamefont\endcsname\relax
  \def\citenamefont#1{#1}\fi
\expandafter\ifx\csname url\endcsname\relax
  \def\url#1{\texttt{#1}}\fi
\expandafter\ifx\csname urlprefix\endcsname\relax\def\urlprefix{URL }\fi
\providecommand{\bibinfo}[2]{#2}
\providecommand{\eprint}[2][]{\url{#2}}

\bibitem[{\citenamefont{Haldane}(1988)}]{haldane-prl88}
\bibinfo{author}{\bibfnamefont{F.~D.~M.} \bibnamefont{Haldane}},
  \bibinfo{journal}{Phys. Rev. Lett.} \textbf{\bibinfo{volume}{61}},
  \bibinfo{pages}{2015} (\bibinfo{year}{1988}).

\bibitem[{\citenamefont{Thouless}(1998)}]{thouless}
\bibinfo{author}{\bibfnamefont{D.~J.} \bibnamefont{Thouless}},
  \emph{\bibinfo{title}{Topological Quantum Numbers in Nonrelativistic
  Physics}} (\bibinfo{publisher}{World Scientific},
  \bibinfo{address}{Singapore}, \bibinfo{year}{1998}).

\bibitem[{\citenamefont{Thouless et~al.}(1982)\citenamefont{Thouless, Kohmoto,
  Nightingale, and den Nijs}}]{tknn}
\bibinfo{author}{\bibfnamefont{D.~J.} \bibnamefont{Thouless}},
  \bibinfo{author}{\bibfnamefont{M.}~\bibnamefont{Kohmoto}},
  \bibinfo{author}{\bibfnamefont{M.~P.} \bibnamefont{Nightingale}},
  \bibnamefont{and} \bibinfo{author}{\bibfnamefont{M.}~\bibnamefont{den Nijs}},
  \bibinfo{journal}{Phys. Rev. Lett.} \textbf{\bibinfo{volume}{49}},
  \bibinfo{pages}{405} (\bibinfo{year}{1982}).

\bibitem[{\citenamefont{Brouder et~al.}(2007)\citenamefont{Brouder, Panati,
  Calandra, Mourougane, and Marzari}}]{brouder-prl07}
\bibinfo{author}{\bibfnamefont{C.}~\bibnamefont{Brouder}},
  \bibinfo{author}{\bibfnamefont{G.}~\bibnamefont{Panati}},
  \bibinfo{author}{\bibfnamefont{M.}~\bibnamefont{Calandra}},
  \bibinfo{author}{\bibfnamefont{C.}~\bibnamefont{Mourougane}},
  \bibnamefont{and} \bibinfo{author}{\bibfnamefont{N.}~\bibnamefont{Marzari}},
  \bibinfo{journal}{Phys. Rev. Lett.} \textbf{\bibinfo{volume}{98}},
  \bibinfo{eid}{046402} (\bibinfo{year}{2007}).

\bibitem[{\citenamefont{Thonhauser and Vanderbilt}(2006)}]{thonhauser-prb06}
\bibinfo{author}{\bibfnamefont{T.}~\bibnamefont{Thonhauser}} \bibnamefont{and}
  \bibinfo{author}{\bibfnamefont{D.}~\bibnamefont{Vanderbilt}},
  \bibinfo{journal}{Phys. Rev. B} \textbf{\bibinfo{volume}{74}},
  \bibinfo{pages}{235111} (\bibinfo{year}{2006}).

\bibitem[{\citenamefont{Resta and Sorella}(1999)}]{resta-prl99}
\bibinfo{author}{\bibfnamefont{R.}~\bibnamefont{Resta}} \bibnamefont{and}
  \bibinfo{author}{\bibfnamefont{S.}~\bibnamefont{Sorella}},
  \bibinfo{journal}{Phys. Rev. Lett.} \textbf{\bibinfo{volume}{82}},
  \bibinfo{pages}{370} (\bibinfo{year}{1999}).

\bibitem[{\citenamefont{Marzari and Vanderbilt}(1997)}]{marzari-prb97}
\bibinfo{author}{\bibfnamefont{N.}~\bibnamefont{Marzari}} \bibnamefont{and}
  \bibinfo{author}{\bibfnamefont{D.}~\bibnamefont{Vanderbilt}},
  \bibinfo{journal}{Phys. Rev. B} \textbf{\bibinfo{volume}{56}},
  \bibinfo{pages}{12847} (\bibinfo{year}{1997}).

\bibitem[{\citenamefont{King-Smith and Vanderbilt}(1993)}]{kingsmith-prb93}
\bibinfo{author}{\bibfnamefont{R.~D.} \bibnamefont{King-Smith}}
  \bibnamefont{and}
  \bibinfo{author}{\bibfnamefont{D.}~\bibnamefont{Vanderbilt}},
  \bibinfo{journal}{Phys. Rev. B} \textbf{\bibinfo{volume}{47}},
  \bibinfo{pages}{1651} (\bibinfo{year}{1993}).

\bibitem[{\citenamefont{Vanderbilt and King-Smith}(1993)}]{vanderbilt-prb93}
\bibinfo{author}{\bibfnamefont{D.}~\bibnamefont{Vanderbilt}} \bibnamefont{and}
  \bibinfo{author}{\bibfnamefont{R.~D.} \bibnamefont{King-Smith}},
  \bibinfo{journal}{Phys. Rev. B} \textbf{\bibinfo{volume}{48}},
  \bibinfo{pages}{4442} (\bibinfo{year}{1993}).

\bibitem[{\citenamefont{Resta and Vanderbilt}(2007)}]{resta-chapter06}
\bibinfo{author}{\bibfnamefont{R.}~\bibnamefont{Resta}} \bibnamefont{and}
  \bibinfo{author}{\bibfnamefont{D.}~\bibnamefont{Vanderbilt}}, in
  \emph{\bibinfo{booktitle}{Modern Ferroelectrics.}}, edited by
  \bibinfo{editor}{\bibfnamefont{C.}~\bibnamefont{Ahn}} \bibnamefont{and}
  \bibinfo{editor}{\bibfnamefont{K.}~\bibnamefont{Rabe}}
  (\bibinfo{publisher}{Springer-Verlag, Berlin}, \bibinfo{year}{2007}), pp.
  \bibinfo{pages}{31--68}.

\bibitem[{\citenamefont{Souza et~al.}(2000)\citenamefont{Souza, Wilkens, and
  Martin}}]{souza-prb00}
\bibinfo{author}{\bibfnamefont{I.}~\bibnamefont{Souza}},
  \bibinfo{author}{\bibfnamefont{T.}~\bibnamefont{Wilkens}}, \bibnamefont{and}
  \bibinfo{author}{\bibfnamefont{R.~M.} \bibnamefont{Martin}},
  \bibinfo{journal}{Phys. Rev. B} \textbf{\bibinfo{volume}{62}},
  \bibinfo{pages}{1666} (\bibinfo{year}{2000}).

\bibitem[{foo()}]{footnote1}
\bibinfo{note}{Note that $P_1\neq0$ even for $\alpha=0$, since $\Delta\neq0$
  allows for an asymmetric population of the edge states.}

\bibitem[{\citenamefont{Sgiarovello et~al.}(2001)\citenamefont{Sgiarovello,
  Peressi, and Resta}}]{sgiarovello}
\bibinfo{author}{\bibfnamefont{C.}~\bibnamefont{Sgiarovello}},
  \bibinfo{author}{\bibfnamefont{M.}~\bibnamefont{Peressi}}, \bibnamefont{and}
  \bibinfo{author}{\bibfnamefont{R.}~\bibnamefont{Resta}},
  \bibinfo{journal}{Phys. Rev. B} \textbf{\bibinfo{volume}{64}},
  \bibinfo{pages}{115202} (\bibinfo{year}{2001}).

\end{thebibliography}

\end{document}